\documentclass[a4paper]{jpconf}

\def\AEF{Faraggi A E}
\def\S{{\cal S}} 
 
\def\W{{\cal W}} 
 
\usepackage{amsmath,amssymb,lmodern}

\def\NPB#1#2#3{#2 {\it Nucl.\ Phys.}\/ {\bf B#1} #3}
\def\PLB#1#2#3{#2 {\it Phys.\ Lett.}\/ {\bf B#1} #3}
\def\PLA#1#2#3{#2 {\it Phys.\ Lett.}\/ {\bf A#1} #3}
\def\PRD#1#2#3{#2 {\it Phys.\ Rev.}\/ {\bf D#1}  #3}
\def\PRL#1#2#3{#2 {\it Phys.\ Rev.\ Lett.}\/ {\bf #1} #3}
\def\PRT#1#2#3{#2 {\it Phys.\ Rep.}\/ {\bf#1} #3}

\def\IJMP#1#2#3{#2 {\it Int.\ J.\ Mod.\ Phys.}\/ {\bf A#1} #3}

\def\EJP#1#2#3{#2 {\it Eur.\ Phys.\ Jour.}\/ {\bf C#1} #3}
\def\JHP#1#2#3{#2 {\it JHEP}\/ {\bf #1} #3}
\def\JCA#1#2#3{#2 {\it JCAP}\/ {\bf #1} #3}

\newcommand{\beq}{\begin{equation}}
\newcommand{\eeq}{\end{equation}}
\newcommand{\beqa}{\begin{eqnarray}}
\newcommand{\beqn}{\begin{eqnarray}}
\newcommand{\eeqn}{\end{eqnarray}}
\newcommand{\eeqa}{\end{eqnarray}}

\usepackage{epsf}
\usepackage{graphicx}
\begin{document}
\title{The M\"obius Symmetry of Quantum Mechanics}

\author{Alon E. Faraggi$^1$ and Marco Matone$^2$}

\address{$^1$ Department of Mathematical Sciences, University of Liverpool, 
Liverpool L69 7ZL, UK}

\address{$^2$ Dipartimento di Fisica, Universit`a di Padova, 
Via Marzolo 8, I--35131 Padova, Italy}

\ead{alon.faraggi@liv.ac.uk; marco.matone@pd.infn.it}

\begin{abstract}

The equivalence postulate approach to quantum mechanics 
aims to formulate quantum mechanics from a fundamental 
geometrical principle. Underlying the formulation there
exists a basic cocycle condition which is invariant under
$D$--dimensional M\"obius transformations with respect 
to the Euclidean or Minkowski metrics. The invariance
under global M\"obius transformations implies that spatial
space is compact. Furthermore, it implies energy quantisation
and undefinability of quantum trajectories without assuming any prior
interpretation of the wave function. 
The approach may be viewed as conventional quantum mechanics
with the caveat that spatial space is compact, as dictated 
by the M\"obius symmetry, with the classical limit corresponding
to the decompactification limit. Correspondingly, there exists
a finite length scale in the formalism and consequently an 
intrinsic regularisation scheme. Evidence for the compactness of
space may exist in the cosmic microwave background radiation. 

\end{abstract}

\section{Introduction}

The mathematical modelling of observational data on the smallest 
and largest distance scales currently rests on two main theories, 
quantum mechanics and general relativity.
On the smallest scales quantum mechanics, through its incarnation
as the Standard Model of Particle Physics, accounts for 
all subatomic data with a high degree of precision. 
The vindication of the Standard Model received
strong support with the observation of a scalar boson
resonance at the Large Hadron Collider (LHC)
with the expected Standard Model 
properties. Higgs studies will dominate
the experimental particle physics program
in the next few decades, {\it i.e.}:
measuring its couplings to the Standard Model fermions and vector bosons; 
measuring its self--couplings; 
and constraining deviation from the Standard Model by 
constraining higher dimensional nonrenormalisable operators 
suppressed by a higher energy scale.

On the largest scales general relativity 
receives ample support from observations 
in celestial, galactical and cosmological data.
In seeking extensions to the two theories,
it is hard to compare the two sets of 
data as the particle physics experiments
rely on billions of controlled events and a thorough 
understanding of the background, whereas the astrophysical 
and, in particular, the cosmological data, rely on a few, or a 
single, event, and poor understanding of the background. 
This distinction is particularly important when seeking to 
synthesise the two theories, and the weight placed on observations
in the two domains. It is also vital in the consideration of
future experimental facilities and the prospect that they
will lead to improved understanding of fundamental physics. 

While quantum mechanics and general relativity are successful
in accounting for observational data in their respective domains, 
their synthesis is nothing but settled.
Furthermore, the two formalisms follow
conceptually distinct approaches. 
The principles of equivalence and covariance with respect to
general diffeomorphism underly general relativity. That is, 
general relativity follows from a fundamental geometrical 
principle. On the other hand, quantum mechanics does not 
follow from a geometrical principle. The main tenant of
the axiomatic formulation of quantum mechanics is the 
probability interpretation of the wave--function. 

Thus, quantum mechanics and general relativity follow
fundamentally distinct approaches. However, 
the two theories are also fundamentally incompatible. 
This incompatibility is most clearly seen in the 
treatment of the vacuum and the vacuum energy. 
In quantum mechanics 
we have to define a vacuum state on which 
the quantum operators operate and create the 
physical states of the Hilbert space. The vacuum 
is the state which is annihilated by all annihilation 
operators. The vacuum should be bounded 
from below. One issue arises in quantum field theories
due to the existence of an infinite number of states and 
the normal ordering ambiguity\footnote{ 
{\it For lucid introduction to quantum field 
theories, see  e.g. } \cite{majore}.}. 
The zero point energy of the vacuum state
therefore leads to an infinite contribution 
to the vacuum energy. 
In particle physics experiment we only measure 
energy differences and absorb additional infinities 
in physical parameters that are measured experimentally. 
If the number of parameters needed to absorb the infinities
is finite the theory is said to be renormalizable. 
Otherwise it is said to be nonrenomalisable. 
The triumph of the Standard Model is precisely because 
it contains a relatively small number of such 
parameters, and is able to account for a 
much larger number of experimental observations. 
This opens the possibility that the Standard Model
is a valid description of the physical data 
not only at a scale which is within reach of 
contemporary experiments, but up to a much larger 
energy scale. It is clear then that the first priority
of forthcoming experiments is to continue to test the
validity of the Standard Model at increasing energy 
scales, by using effective field theory approach to 
parameterise possible deviations. Furthermore, 
the set of parameters associated with the Higgs 
sector are particularly ripe for experimental picking. 
Gravity on the other hand is known to be nonrenomalizable. 
Furthermore, gravitational measurements are sensitive to the 
absolute energy scale. Observations dictate that the 
vacuum energy is miniscule compared to what we would 
expect from particle physics. The dichotomy 
between the two theories motivates much of the 
contemporary research in theoretical particle physics. 
Different approaches are pursued to develop a consistent
theory that synthesises quantum mechanics and gravity.
These include: 
effective field theories \cite{eft};
Euclidean quantum gravity \cite{eqg};
asymptotic safety \cite{asafety}; 
causal dynamical triangulation \cite{cdt};
twistor theory \cite{twistortheory};
noncommutative geometry \cite{ncg};
loop quantum gravity \cite{lqg};
string theory \cite{stheory}. 
What are we to learn from this very partial list?
First, we note that all these approaches aim at 
quantising general relativity, {\it i.e.} 
quantising spacetime. 
Second, non of these theories produces
an unequivocal signature that has been confirmed 
experimentally, and in that respect, all these 
attempts should be regarded on equal footing. 
Nevertheless, the most developed effort is 
undoubtedly that of string theory.
The main successes of string theory is that: 
1. it provides a viable perturbative approach to quantum gravity;
2. it produce the gauge and matter structures that underly 
the Standard Model. As such it provides
a framework for the construction of phenomenologically realistic 
models, and is therefore relevant for experimental observations. 
In that respect, by unifying the gauge and matter
sectors with gravity, string theory provides a framework 
to study how the parameters of the Standard Model may arise 
from a more fundamental theory, and goes beyond the 
field theoretic Grand Unified Theories. Moreover, 
string theory accommodates consistently one
of the most intriguing properties of the observed 
particles, that of chirality and with it parity 
violation. It achieves that by its most important 
property, modular invariance, which underlies the 
anomaly cancellation. By achieving that string theory
demonstrated its potential relevance for experimental data, 
though we may still be a long way off before we can 
describe in a detailed and rigorous way how this relevance 
is realised in the material world. The state of the 
art in this endeavour is the derivation of the Minimal 
Supersymmetric Standard Model from heterotic--string
theory \cite{mshsm}. These models demonstrate how the detailed structures
of the Standard Model may arise from string theory. 
In particular, the models enabled the calculation 
of the Yukawa couplings of the top quark, bottom quark
and tau lepton in terms of the gauge coupling at the 
unification scale. Assuming consistency with the 
low scale gauge data then enables extrapolation of the
Yukawa coupling to the low scale and prediction of the 
top quark mass of order $O(175 GeV)$ \cite{top} in the vicinity of 
the observed value. These results unambiguously demonstrate 
the potential relevance of string theory for low scale
experimental data. Furthermore, the string consistency 
conditions dictate that additional extra degrees of 
freedom beyond the Standard Model, 
are needed to obtain a consistent theory of 
perturbative quantum gravity, which is an unambiguous 
prediction of the theory. These may be interpreted as 
extra spacetime dimensions, and/or as additional gauge 
symmetries beyond those observed in the Standard Model.
It is a secondary question whether we possess the 
technological tools to detect experimentally 
these additional degrees of freedom.
The myriad of string solutions which 
may in principle be compatible with the low energy data 
places an additional hindrance on extracting unambiguous 
experimental signatures from string theory. 
Given the hierarchical separation between the 
strength of the gravitational and gauge couplings
it may be a long while before such an unambiguous 
correlation may be extracted. In that respect progress
in string theory is likely to be incremental and 
slow. Another limitation is that our current 
understanding of string theory is limited to
asymptotic stable string solutions, and we lack a
good understanding of it as a dynamical theory. 
This hinders the development of cosmological string scenarios
and of a possible dynamical vacuum selection mechanism in string theory.
One then has to resort to analysis in the effective low energy field theory
limit of the string vacua, which is deficient because it 
misses the massive string spectrum and the possible roll that
it may play in the string vacuum selection mechanism. 
However, as long as the low scale data
does not indicate departure from perturbative 
Standard Model parameterisation of the experimental 
observations, string theory continues to provide 
the most detailed framework to calculate the Standard 
Model parameters from a more fundamental theory.

String theory therefore provides a viable perturbative 
framework to explore how the Standard Model parameters may
arise from a fundamental theory and to develop a phenomenological 
approach to quantum gravity. However, string theory does not 
provide a fundamental physical principle, akin to the equivalence
principle of general relativity, that we may use
as the basic hypothesis, and formulate quantum gravity 
rigoursely staring from that hypothesis. For that 
we may need an entirely new approach. In that respect 
string theory may be viewed as an effective theory, 
perhaps in a similar spirit to the view of effective
quantum field theories as effective theories. 
However, string theory may hint at a possible 
basic hypothesis from which string theory may be 
derived as an effective limit. 

An important property of string theory is the relation that it exhibits 
between different vacua by perturbative and non--perturbative
transformations, and
the existence of self--dual states under the 
duality transformation. We may envision that the self--dual 
states, which represent enhanced symmetry points in the 
space of vacua, may play a role in the string vacuum 
selection mechanism. We may also imagine that the 
duality structures underlying string theory
may provide a hint at the basic hypothesis 
from whence string theory may be derived.
In that context, however, 
it is noted that the string dualities correspond to 
specific dualities between concrete vacua and detailed 
mathematical structures. 
In that respect they provide examples that possess
too much structure and are not sufficiently 
rudimentary to provide a basic physical hypothesis.
It therefore does not provide a sufficiently basic
physical hypothesis. 
Rather, we may consider that the perturbative and 
non--perturbative dualities exhibited in string theory
may all be regarded as dualities in an extended phase space.

Phase space represents the key transition from the classical 
to the modern description of physical experiments.
In that respect we need to establish 
what is the usefulness of a 
physical theory. 
We should first prioritise physics 
as an experimental science. 
We may define physics as ``mathematical 
modelling of experimental observations''.
The aim of a mathematical model is to predict the 
outcome of experiments. Starting with some initial 
conditions, which are fed into the mathematical 
model, the predicted outcomes are calculated, 
and are then confronted with experiments. 
Physics is a practical field and an 
accepted mathematical model is the one
most successful in accounting for a 
wide range of experimental observations. 
From that perspective, as our technological 
tools develop, our capacity to make experimental
observations advances with time. Consequently, 
our mathematical modelling evolves with time 
to accommodate the expanding body of experimental data. 
This process is traditionally termed as reductionism. 
Namely, as experimental tools evolve we 
can resolve more refined physical distances. 
From celestial in the Galilean era to the sub-nuclear 
in the LHC era. The process of adapting our 
mathematical models to the increased body 
of experimental data may be labelled as ``unification'', 
and is likewise a constant theme in the mathematical 
modelling of experimental data. Thus, for example, 
Newton unified terrestrial and celestial observations in Newtonian mechanics;
Maxwell combined the electric and magnetic forces into electromagnetism; 
Einstein synthesised mechanics and electromagnetism in special relativity; 
Glashow, Salam and Weinberg meshed quantum electrodynamics with Marshak and
Sudarshan's vector minus axial--vector theory of the weak interactions 
into the electroweak model.
The Standard Model that describe all the subatomic 
interactions as gauge theories may also be regarded as unifying
the subatomic interactions under a single physical principle.

The list above represent the steps in the mathematical modelling
of experimental data in which diverse observations are described 
in a common framework. The next stops on this route are the
ones for which we do not yet have experimental evidence. They may
therefore not necessarily be realised in nature.
These include 
Grand Unified Theories which 
unify the electroweak and strong interactions 
in a simple or semi--simple group. The main prediction of 
Grand Unified Theories is proton decay. Global supersymmetry 
which combines fermions and bosons in common multiplets. 
The generic prediction of the simplest realisations of 
supersymmetry is the existence of superpartners. However, 
non--linear realisations of supersymmetry may exist, in
which this will not be the case. 
Local supersymmetry implies the existence of a spin 3/2 
particle, and consequently the existence of a spin 2 particle. 
Local supersymmetry therefore unifies gravity with the gauge 
interactions, albeit in a nonrenormalisable theory. 

In Newtonian mechanics the mathematical modelling of the 
physical systems uses the position coordinates of a 
particle in space and their derivative with respect 
to time, {\it i.e.} the velocities. Given the 
position and velocity at some initial time $t=t_i$, 
and given some force field ${\vec F}({\vec x})$, 
the evolution of the position and velocity coordinates
is determined by Newton's equation of motion.
A fully equivalent classical description is 
provided by exchanging the velocities of the 
particles with their momenta. The transformation is
from the configuration space to the phase space. 
An equivalent representation of Newton's equations of motion
in classical mechanics is given given in terms of the
Euler--Lagrange equations of motion, which are derived 
by defining a Lagrangian function in configuration space. 
By transforming from the configuration space to the phase space,
we transform the Lagrangian to the Hamiltonian function in 
phase space by a Legendre transformation. 
In the process we transform the second order 
Euler--Lagrange equations to the set of first order
Hamilton equations of motion.
The price that we have to pay is the doubling 
of the number of equations. Namely for every 
second equation we have two first order equations.
The Hamilton equations of motion are 
invariant, up to a sign, under exchange of 
coordinates and momenta.
Similar to the Newton equations of motion, 
the Hamilton equations of motion account 
for the time evolution of the phase space 
variables. 

The relevance of a physical theory is established by
confronting a mathematical model with experimental
data. Typically the experimental setup and the 
variables of the mathematical model may evolve with 
time. Therefore, an important set of variables in
the mathematical model are those that do not change 
as the physical system evolves. These are the 
constants of the motions and are related 
to the symmetries of the physical systems. 

\section{The Hamilton--Jacobi Theory}

A general method to solve a physical problem in 
classical mechanics is given by the Hamilton--Jacobi 
formalism. In the classical Hamilton--Jacobi formalism
the solution of the physical problem 
is obtained by transforming the Hamiltonian
from one set of phase--space variables to a new 
set of phase--space variables, such that 
Hamiltonian is mapped to a trivial Hamiltonian. 
We may refer to these transformations as trivialising 
transformations. 
Consequently, the new phase--space variables are 
constants of the motion, {\it i.e.}, 
\begin{equation}
H(q,p)~~\longrightarrow~~ { K}({ Q},{P})\equiv0~~~~~~~~
\Longrightarrow~~~~~\dot{  Q}={\partial{K}\over
{\partial{P}}}\equiv 0~,~\dot{P}=
-{\partial {K}\over{\partial Q}}\equiv 0. ~~~~~~~~~
\label{hjformalism}
\end{equation}
The solution to this problem is given by the Classical
Hamilton--Jacobi Equation (CHJE), which in the stationary case 
is given by
\beq
{1\over{2m}}\left({{\partial{ S}_0}\over
{\partial q}}\right)^2~~+~~V(q)~~-~~E~~\equiv~~
{1\over{2m}}\left({{\partial{ S}_0}\over
{\partial q}}\right)^2~~+W(q)~~=~~0. 
\label{hje}
\eeq
The phase space variables are taken to be independent 
and their functional dependence is only extracted 
from the solution of the Hamilton--Jacobi equation
via the relation
\begin{equation}
p={{\partial S(q)}\over {\partial q}}, 
\label{pvsq}
\end{equation}
where the generating function $S(q)$ is Hamilton's principal 
function. The key property of quantum mechanics is that 
the phase space variables are not independent. Namely, 
represented as quantum operator they do no commute. 
We may therefore envision posing a similar problem to the
Hamilton--Jacobi procedure, but one in which the 
phase--space variables are not treated as independent variables, 
but are rather related by (\ref{pvsq}).

\section{The cocycle condition}

We first consider the stationary problem 
in the one dimensional case \cite{fm2, fmut, fmtqt, fmpl, fmijmp}. 
This reveals the 
general properties of the formalism and 
paves the way for the general cases. 
We therefore assume that there always exist 
coordinate transformations such that $H\rightarrow K\equiv 0$, 
{\it i.e.} such that in the new system both the potential energy and
the kinetic energy vanish. More generally, impose that there 
exist coordinate transformations such that in the 
transformed system $W(Q)=V(Q)-E\equiv 0$. 
We consider the transformations on
\begin{equation}
(~q~,~S_0(q)~,~p~=~{{\partial S_0}\over{\partial q}})~~\longrightarrow~~
(~{\tilde q}~,~{\tilde S}_0({\tilde q})~,~{\tilde p}~=~
{{\partial{\tilde S}_0}\over{\partial{\tilde q}}}~), 
\label{onedtransformations}
\end{equation}
such that 
$
W(q)~~\longrightarrow~~{\tilde W}({\tilde q})~=~0,~
$
exist for all $W(q)$. We refer to this proposition as the
``equivalence postulate of quantum mechanics''. 
It implies that all physical systems, labelled 
by a potential function $W(q)$, can be connected 
by coordinate transformations. More narrowly, 
we may regard it as adaptation of the 
classical Hamilton--Jacobi formalism
to quantum mechanics. Irrespective of the 
concrete interpretation, it reveals the
M\"obius symmetry that underlies quantum mechanics.
The equivalence postulate implies that the Hamilton--Jacobi (HJ)
equation has to be covariant under coordinate transformations. 
However, from eq. (\ref{hje}) it is seen that the first term
transforms as a quadratic differential, whereas the 
potential function $W(q)$, in general, does not.
Furthermore, the state $W(q)\equiv 0$ is fixed
under the transformations $q\rightarrow {\tilde q}(q)$. 
We therefore assume that the HJ equation is modified
by adding a yet to be determined function $Q(q)$. 
The modified HJ equation then takes the form 
\beq
~{1\over{2m}}\left({{\partial{ S}_0}\over {\partial q}}
\right)^2~~+~~W(q)~~+~~Q(q)~~=~~0, 
\label{mhje}
\eeq
where under the transformations $q\rightarrow {\tilde q}(q)$
the two functions $W(q)$ and $Q(q)$ transform as 
\beqn
{\tilde W}({\tilde q}) &=& 
\left({{\partial {\tilde q}}\over{\partial q}}\right)^{-2}W(q)+
({\tilde q};q),\nonumber\\
{\tilde Q}({\tilde q}) &=& 
\left({{\partial {\tilde q}}\over{\partial q}}\right)^{-2}Q(q)-
({\tilde q};q),\nonumber
\eeqn
with ${\tilde S}_0({\tilde q})~=~S_0(q)$. It is seen 
that each of the functions $W(q)$ and $Q(q)$ transforms 
as a quadratic differential up to an additive term and that
the combination $W(q)+Q(q)$ transforms as a quadratic 
differential. Starting with the trivial 
state $W^0(q^0)\equiv 0$, all other physical states arise
from the additive inhomogeneous term as $W(q)\equiv (q;q^0)$.
Furthermore, considering the transformations 
$q^a\rightarrow q^b \rightarrow q^c$ and $q^a\rightarrow q^c$ 
and the induced transformations 
$W^a(q^a)\rightarrow W^b(q^b) \rightarrow W^c(q^c)$
and $W^a(q^a)\rightarrow W^c(q^c)$
gives rise to a cocycle condition 
on the inhomogeneous term given by
\beq
(q^a; q^c)~=~\left({{\partial q^b}\over {\partial q^c}}\right)^2
\left[~(q^a; q^b)~-~(q^c; q^b)\right].
\label{cocyclecon}
\eeq
It can then be shown that the cocycle condition is 
invariant under M\"obius transformations $\gamma(q^a)$. 
In the one dimensional case the M\"obius 
symmetry uniquely fixes the functional form
of the inhomogeneous term to be given by 
the Schwarzian derivative, {\it i.e.}
$
(q^a;q^c)~~\sim~~\left\{q^a; q^c\right\},
$
where the Schwarzian derivative is defined by 
\beq
\{f(q),q\}= {f^{\prime\prime\prime}\over{f^\prime}}- 
{3\over2}\left({f^{\prime\prime}\over{f^\prime}}\right)^2.
\label{schwarzian}
\eeq

\section{The Quantum Hamilton--Jacobi Equation}\label{qhje}
 
Considering the Schwarzian identity 
\beq
\left({{\partial S_0}%
\over{\partial q}}\right)^2={\beta^2\over{2}}\left(\{{\rm e}^{{i2S_0}\over%
\beta};q\}-\{S_0;q\}\right),
\label{sidentity}
\eeq
we note that the quadratic differential on the left--hand 
side of the equation is written as the difference of two 
Schwarzian derivatives. As we will see,
the cocycle condition eq. (\ref{cocyclecon})
and the Schwarzian identity eq. (\ref{sidentity})
lay down the key ingredients for the generalisations
to higher dimensions. Equally fundamental is the
invariance of the cocycle condition and of the Schwarzian 
derivative under M\"obius transformations. It is proven 
rigoursely \cite{bfm} that the corresponding 
cocycle condition in $D$--dimensions is invariant 
under $D$--dimensional M\"obius transformations,
with respect to the Euclidean or Minkowski metrics. 
Similarly, the generalisation of the Schwarzian
identity eq. (\ref{sidentity}) entails writing the 
quadratic differential on the left--hand side 
as a generalised identity on the right--hand side. 
Making the identifications
\begin{eqnarray}
W(q) & = & -{\beta^2\over{4m}}\{{\rm e}^{{i2S_0}\over\beta};q\}=V(q)-E, \nonumber\\
Q(q) & = & {\beta^2\over{4m}}\{S_0;q\}, \nonumber
\end{eqnarray}
the modified Hamilton--Jacobi equation becomes, 
\beq
{1\over{2m}}\left({{\partial S_0}\over{\partial q}}\right)^2+%
V(q)-E+{\beta^2\over{4m}}\{S_0;q\}=0. 
\label{mshje}
\eeq
The key property of quantum mechanics is gleaned from eq (\ref{mshje}). 
In the CSHJE admits the solution $S_0=constant$ for the state
$W^0(q^0)\equiv 0$. The
Quantum Stationary Hamilton--Jacobi Equation (QSHJE), eq. (\ref{mshje}), 
admits the solutions, $$S_0=\pm{\beta\over2}\log q\ne A q + B,$$
where $A$ and $B$ are constants. Thus, in quantum mechanics
$S_0$ is never a constant, and more generally is never a 
linear function of $q$. This is a key property 
of quantum mechanics in this formalism, and is 
intimately related to consistency of phase--space duality. 
Furthermore, from the properties of the Schwarzian derivative
we know that the function $W(q)=V(q)-E$
is a potential of a second order differential equation given by
\beq
\left(-{\beta^2\over{2m}}{\partial^2\over{\partial q^2}}+V(q)-E\right)%
\Psi(q)=0, 
\label{secondorderdiffeq}
\eeq
which we identify as the Schr\"odinger equation, and 
${\beta~=~\hbar}$ as the covariantising parameter of the 
Hamilton--Jacobi equation. The general solutions of 
(\ref{secondorderdiffeq}) and (\ref{mshje}) are given by 
\beq
\Psi(q)=\psi_1+\psi_2=
{1\over{\sqrt{S_0^\prime}}}\left(A{\rm e}^{+{i\over\hbar}S_0}+
B{\rm e}^{-{i\over\hbar}S_0}\right)
\label{wavefunction}
\eeq
and
\beq
{\rm e}^{+{{i2S_0}\over\hbar}}~=~{\rm e}^{i\alpha}~%
{{w+i{\bar\ell}}\over{w-i\ell}}
\label{qhjesolution}
\eeq
where $w~=~{\psi_1/\psi_2} $, 
$\ell~=~\ell_1~+~i~\ell_2$, and
$\ell_1~\ne~0$, $\alpha~\in~ R$.
The condition $\ell_1\ne 0$ is synonymous to the 
condition that $S_0\ne constant$.

It is noted that the Schr\"odinger equation serves as
a tool to solve the Quantum Hamilton--Jacobi Equation (QHJE). 
Hence, the more fundamental equation is the QHJE, 
and captures the symmetry properties that underly 
quantum mechanics. In that respect the solutions of the
Schr\"odinger equation, eq. (\ref{wavefunction}),
facilitate the solution of the QHJE via eq. (\ref{qhjesolution}).

It is further seen that consistency of the 
formalism necessitates that both solutions of the 
Schr\"odinger equation are used. This is evident
in eq. (\ref{qhjesolution}) from which we see that 
the ratio of the two solutions of the Schr\"odinger 
equation is used to extract the solution for the 
the QHJE. It is a reflection
of the M\"obius symmetry that underlies the formalism, 
and has implication on the global geometry that underlies 
quantum mechanics and quantum gravity. The basic point is that 
the M\"obius symmetry includes a symmetry under inversions. 
In the case of $D$--dimensional spacetime this can be implemented 
via inversions with respect to the unit sphere. More generally, 
the global geometry, whatever it may be, has to be invariant
under the global M\"obius transformations. In the case of 
spatial space, {\it i.e.} with Euclidean signature of the 
metric, the implication is that spatial space must be compact. 
Otherwise the invariance under the M\"obius symmetry cannot 
be applied consistently. Therefore, at the basic level what we
face in comparison to conventional quantum mechanics, is 
a question of boundary conditions. Namely, in the 
case of bound physical systems in conventional quantum mechanics
the solutions of the Schr\"odinger equation include a square integrable 
solution and a solution that diverges at infinity. Therefore, the physical
solution is retained, whereas the solution that diverges at infinity 
is non--physical and hence discarded. In essence, however, 
it is a question of boundary conditions. Namely, if space is infinite 
then the diverging solution may be discarded. However, if space 
is compact, as implied by the M\"obius symmetry, then the diverging 
solution cannot be discarded and must be retained. The wave function 
is necessarily a combination of the two solutions, albeit the coefficient 
of the diverging solution may be unobservationally small. In a sense 
therefore the equivalence postulate approach may be regarded as conventional 
quantum mechanics plus the condition that spatial space is compact, as dictated
by the M\"obius symmetry that underlies quantum mechanics. Consistency of the
formalism therefore entails that the transformation
$W(q)=V(q)-E\longrightarrow{\tilde W}({\tilde q})=0$
always exists, and is given by $q\rightarrow {\tilde q}=\psi_1/\psi_2$. 
Applying the transformation then entails that 
\beq
\left(-{\beta^2\over{2m}}~{\partial^2\over{\partial q^2}}~+~%
V(q)~-~E\right)\psi(q)~=~0~~\rightarrow~~%
{\partial^2\over{\partial{\tilde q}^2}}%
{\tilde\psi}({\tilde q})~=~0
\label{trivialse}
\eeq
where 
$${\tilde\psi}({\tilde q})~=~\left({{dq}%
\over{d{\tilde q}}}\right)^{-{1\over2}}{\psi(q)}.$$

\section{Energy quantisation}

In conventional quantum mechanics the probability interpretation
implies that the wave function and its derivative are continuous
and is square integrable. Consequently, for bound states the 
the energy level are quantised. The question is therefore,
how is it replicated in the equivalence postulate approach?
As the trivialising transformation is given by
\beq 
q^0~=~w~=~{\psi_1\over\psi_2}~=~{\psi^D\over\psi}
\eeq
and is a solution of 
\beq
\{w,q\}~=~-{{4m}\over\hbar^2}(V(q)-E)
\eeq
we have that 
$
w\ne const;~w\in C^2(R)~ and~ w^{\prime\prime}~%
\hbox{differentiable on}~R.~
$
From the properties of the Schwarzian derivative 
under inversions it follows that 
$
\{w,q^{-1}\}~=~q^4\{w,q\}.
$
Hence, the consistency conditions on the trivialising transformation
must be imposed not merely on the real line, but on the extended 
real line, {\it i.e.} on the real line plus the point at infinity. 
That is, 
$$w~\ne~ const~~;~~w~\in~ C^2({\hat R})~~ and~~ %
w^{\prime\prime}~\hbox{differentiable on}~{\hat R},~~~\hbox{where}~~~
{\hat R}~=~R~\cup~\{\infty\}.
$$
Consequently, the equivalence postulate implies the continuity 
of the two solutions of the Schr\"odinger equation and 
their derivatives. Furthermore, a general theorem that states
that if the potential function $W(q)$ is bounding in some 
interval, then the ratio $w=\psi_1/\psi_2$ is continuous on the 
extended real line if and only if the Schr\"odinger equation
admits a square integrable solution, is proven rigoursely in
\cite{fmtqt,fmijmp,fmmt}. Rather than go through the theorem it is instructive
to consider the simple problem of a particle in a potential well with
\beq
V(q)=\left\{\begin{array}{ll}0,&|q|\leq L,\\ V_0,&|q|> L.\end{array}\right.
\label{V0}\eeq
Setting $k={\sqrt{2mE}/\hbar}$ and $K={\sqrt{2m(V_0-E)}/\hbar}$, 
the general solutions in and outside the potential well are given by
\begin{eqnarray}
{|q|~\le~ L}~~~~~~~~~~~ & &
            \Psi_1^1~=~\cos{kq}~~~~~~~~~~\Psi_2^1~=~\sin{kq}~~~~~\nonumber\\
    {q~~>~L~~~~~~~~~~~} & & 
            \Psi_1^2~=~{\rm e}^{-Kq}~~~~~~~~~~~\Psi_2^2~=~{\rm e}^{Kq}~~~\nonumber
\end{eqnarray}
The solutions at  $q<-L$ are fixed by parity. 
For $q>L$ we can choose the solution to be $\Psi^2_1$ or $\Psi^2_2$ 
or a linear combination. Continuity across the boundary at $q=L$ implies
that $\Psi_i^1(L)=\Psi^2_j(L)$ and
$\partial_q\Psi^1_i(L)=\partial_q\Psi^2_j(L)$. Denoting 
such solutions as $(i,j)$, in the $(1,1)$ case imposing continuity on the 
solution and its derivative yields the quantisation condition
$k~\tan kL~=~K$. In this case
\beq
\psi=\left\{
\begin{array}{ll}\cos(kL)\exp[K(q+L)],&q<-L,\\ 
                             \cos(kq) ,&|q|\leq L,\\ 
                             \cos(kL)\exp[-K(q-L)],&q>L,
\end{array}\right.
\label{psi11}
\eeq
and a linearly independent solution is given by
\beq
\psi^D=[2k\sin(kL)]^{-1}\cdot\left\{
\begin{array}{ll}\cos(2kL)\exp[K (q+L)]-\exp[-K(q+L)],&q<-L,\\ 
                                          2\sin(kL)\sin(kq),&|q|\leq L,\\
              \exp[K(q-L)]-\cos(2kL)\exp[-K(q-L)],&q>L.
\end{array}\right.
\label{psid11}
\eeq
The trivialising map $w=\psi^D/\psi$ associated with the $(1,1)$ solution
is therefore given by
\beq
{\psi^D\over\psi}=[k\sin(2kL)]^{-1}\cdot\left\{
\begin{array}{ll}\cos(2kL)- \exp[-2K(q+L)],&q<-L,\\ 
\sin(2kL)\tan(kq),&|q|\leq L,\\
\exp[2K(q-L)]-\cos(2kL),&q>L.\end{array}\right.
\label{psidoverpsi11}
\eeq
In this case 
$\lim_{q\longrightarrow\pm\infty}{\psi^D/\psi}=\pm\infty.$
Hence, in the $(1,1)$ case the trivialising map is 
continuous on ${\hat R}$ as required by consistency 
of the equivalence postulate. The solutions imposed 
by the continuity conditions $k~\tan kL~=~K$ are 
therefore physical energy levels. Next, considering 
the case $(1,2)$ the constraint that $\Psi,~\Psi^\prime$
are continuous across at $q=L$ imposes the condition
$k~\tan(kL)~=~-K$. Applying similar analysis to the 
$(1,1)$ case, the trivialising transformation $w$ is give by
\beq
{\psi^D\over\psi}=[k\sin(2kL)]^{-1}\cdot\left\{
\begin{array}{ll}\cos(2kL)-\exp[2 K(q+L)],&q<-L,\\ 
\sin(2kL)\tan(kq),&|q|\leq L,\\ 
\exp[-2 K(q-L)]-\cos(2k L),&q>L,\end{array}\right.
\label{psidoverpsi21}
\eeq
whose asymptotic behaviour is
\beq
\lim_{q\longrightarrow\pm\infty}{\psi^D\over\psi}=\mp k^{-1}\cot(2kL).
\label{watinfty}
\eeq
The only possibility of having $w(-\infty)=w(+\infty)$
is that $k^{-1}\cot(2kL)=0$. However, this is not compatible
with the condition $k~\tan(kL)~=~-K$. Hence, we have
$
w(-\infty)\ne w(+\infty).~
$
It follows that the energy eigenvalues associated with this solution 
are not consistent with the equivalence postulate and are therefore not physical.
Therefore, the same physical eigenstates that are selected in convential 
quantum mechanics by the probability interpretation of the wave function, 
are selected in the equivalence postulate approach by mathematical consistency. 
Essentially respecting the M\"obius symmetry that underlies quantum mechanics. 
We further note that the requirement that the trivialising transformation 
is continuous on the extended real line, amounts to the requirement that
the real line is compact. Therefore, energy quantisation and square integrability
arises from the consistency of the equivalence postulate and the compactness
of space, which is mandated by the M\"obius symmetry underlying quantum mechanics.
However, the probabilistic nature of quantum mechanics, rather than a 
deterministic parameterisation of a particle propagation is 
indicated by many experiments. A viable question is whether such 
parameterisation is consistent with the quantum Hamilton--Jacobi formalism
and the M\"obius symmetry that underlies it. 

\section{Time parameterisation}

There are two primary approaches to define parameterisation of 
trajectories in quantum mechanics. The first is Bohmian mechanics 
\cite{holland,wyatt} in which time parameterisation is defined
by identifying the conjugate momentum with the mechanical momentum, 
{\it i.e.} $p=\partial_q S= m {\dot q}$, where $S$ is the solution 
of the QHJE. The second is Floyd's definition of time parameterisation 
by using Jacobi's theorem, 
$
t={\partial_E S_0^{QM}},
$
where $S_0^{QM}$ is the solution of the QSHJE. In classical mechanics 
the two definitions are compatible.  Namely, setting
$
p=\partial_q{\cal S}_0^{\rm cl}=m{\dot q}~
$
gives
\beq
t-t_0=m\int_{q_0}^q{{dx}\over {\partial_x {\cal S}}_0^{\rm cl}}=
\int_{q_0}^q dx {{\partial~}\over{\partial E}}
{\partial_x {\cal S}}_0^{\rm cl}=
{{\partial {\cal S}_0^{\rm cl}}\over{\partial E}}, 
\eeq
and by inverting we can solve the equation of motion for $q=q(t)$. 
However, in quantum mechanics the two definitions are not 
compatible as 
\beq
t-t_0={{\partial S_0^{\rm qm}}\over{\partial E}} =
{{\partial~}\over{\partial E}} \int_{q_0}^q dx 
{\partial_x S}_0^{\rm qm}=
\left(m\over 2\right)
\int_{q_0}^q dx {{1-\partial_E Q}\over {\left(E-V-Q\right)^{1/2}}}.
\eeq
and the mechanical momentum is given by
\beq
m{{dq}\over {dt}}= m \left({{dt}\over{dq}}\right)^{-1}=
{{\partial_q S^{\rm qm}_0}\over{\left(1-\partial_E{\cal V}\right)}}
\ne\partial_qS_0^{\rm qm},
\eeq
where $\cal V$ denotes the combined potential 
${\cal V}=V(q)+Q(q)$. Thus in quantum mechanics the definition of 
time by using Jacobi's theorem does not coincide with the 
Bohmian definition of time by identifying the conjugate 
momentum with the mechanical momentum. 
Furthermore, the Bohmian time definition is not compatible with
the M\"obius symmetry that underlies quantum mechanics, 
and the compactness of space. In Bohmian 
mechanics the wave function is set as 
\beq
\psi({\vec q},t)=R(q){\rm e}^{iS/\hbar},
\label{bohmpsiq}
\eeq
where $R(q)$ and $S(q)$ are the two real functions of the 
QHJE, and $\psi(q)$ is a solution of the Schr\"odinger equation.
The conjugate momentum is given by 
$$p={\hbar}{\rm Im}{{\nabla\psi}\over\psi},$$
which is used to define trajectories by setting $p=m{\dot q}$.
However, the M\"obius and compactness of space 
dictate that the wave function cannot be identified 
by (\ref{bohmpsiq}) but must be a linear combination 
of the two solutions of the Schr\"odinger equation, 
\beq
\psi= R(q) \left( A {\rm e}^{{i\over \hbar} S} + 
B {\rm e}^{-{i\over \hbar} S}\right). 
\label{onedwavefunction}
\eeq
Consequently, in this case
$$\nabla S\ne {\hbar}{\rm Im}{{\nabla\psi}\over\psi}$$
and the Bohmian definition of trajectories is invalid \cite{fmtpqt}.

Floyd proposed to define time by using Jacobi's theorem
\cite{floyd}, {\it i.e.}
\beq 
t-t_0= {{\partial S_0^{\rm qm}}\over{\partial E}} , 
\label{floydproposal}
\eeq 
which provides a time parameterisation of the trajectories
by inverting $t(q)\rightarrow q(t)$. However, if space is 
compact, as dictated by the M\"obius symmetry that underlies
the QHJE, then the energy levels are always quantised. 
Therefore, differentiation with respect to energy is ill defined, 
and the definition of time parameterisation of trajectories 
by using Jacobi's theorem is inconsistent in the Quantum 
Hamilton--Jacobi formalism. In quantum mechanics time can only be
used as a classical background parameter. The trajectory 
representation can only be used in a semi--classical 
approximation and in that context provides a useful tool
to study different physical systems \cite{wyatt}.

\section{Generalisations} 

The discussion so far focussed on the one dimensional stationary case. 
This led to the cocycle condition, eq. (\ref{cocyclecon}),
and the Schwarzian identity, eq. (\ref{sidentity}), and
their invariance under M\""obius transformations. These are 
the key ingredients that pave the way for the generalisations
to higher dimensions in Euclidean or Minkowski space, {\it i.e.}
for the non-relativistic Schr\"odinger equation or
the relativistic Klein--Gordon equation \cite{bfm}. Furthermore, 
these generalisations are invariant under M\"obius 
transformations in those spaces \cite{bfm}. 
Thus, the M\"obius symmetry captures the global symmetry 
structure that underlies quantum mechanics and 
dictates that spatial space is compact.
Considering the transformations between two $D$--dimensional 
coordinate systems given by 
\beq
q\rightarrow q^v = v(q)
\label{qtovq}
\eeq
and the induced transformations on the conjugate momenta
\beq
p_k= {{\partial S_0}\over {\partial q_k}}, 
\label{pkmomenta}
\eeq
with $S_0^v(q^v)=S_0(q)$. Consequently, 
\beq
p_k\rightarrow p_k^v=\sum_{i=1}^D J_{ki}p_i
\label{pkv}
\eeq
and the Jacobian matrix $J$ is given by
\beq
J_{ki}={{\partial q_i}\over {\partial q_j^v}}.
\label{jacobian}
\eeq
Adopting the notation 
\beq
(p^v|p)={{\sum_k (p_k^v)^2}\over{\sum_kp_k^2}}={{p^tJ^tJp}\over {p^tp}}. 
\label{pvp}
\eeq
 the cocycle condition takes the form 
\beq
(q^a;q^c)=(p^c|p^b)\left[(q^a;q^b)-(q^c;q^b)\right]. 
\label{cocycleinEspace}
\eeq
The cocycle condition, eq. (\ref{cocycleinEspace})
is invariant under $D$--dimensional M\"obius transformations, 
which include translations, rotations, dilatations, and reflections 
with respect to the unit sphere \cite{bfm}. In the case of Minkowski 
spacetime the invariance is with respect to the $D+1$--dimensional
conformal group, where $q\equiv(ct, q_1, \cdots, q_D)$. 
The second key ingredient of the one dimensional 
formalism is the Schwarzian identity, equation ({\ref{sidentity}).
The generalisation of this identity provides the key for the 
extension of the formalism to higher dimensions 
in Euclidean and Minkowski spaces.
In the non--relativistic case the identity is given by
\beq 
\alpha^2(\nabla S_0)^2=
{\Delta(R{\rm e}^{\alpha S_0})\over R{\rm e}^{\alpha S_0}}-
{\Delta R\over R}-{\alpha\over R^2}\nabla\cdot(R^2\nabla S_0), 
\label{ddidentity}
\eeq
which holds for any two real functions 
$R$ and $S_0$ and any constant $\alpha$. Similarly, in the 
relativistic case the generalisation is given by \cite{bfm}
\beq
\alpha^2(\partial\S)^2={\Box(Re^{\alpha\S})\over Re^{\alpha\S}} 
-{\Box R\over R}-{\alpha\over R^2}\partial \cdot (R^2\partial\S), 
\label{ridentity}
\eeq
whereas if we include minimal coupling in the form of a vector
potential the identity takes the form
\beq
\alpha^2 (\partial { S}-eA) \cdot (\partial { S}-eA) =%
{D^2(R e^{\alpha S}) \over R e^{\alpha S}}%
- {\partial^2 R \over R} - {\alpha\over R^2}%
\partial \cdot\left(R^2 (\partial  S-eA)\right),
\label{idwithvp}
\eeq
where $D^\mu~=~\partial^\mu-\alpha e A^\mu$. 
Similarly to the one dimensional case, the higher dimensional 
cases are related to the conventional quantum mechanical 
equations. For example, in the case of eq. (\ref{ridentity}), setting
$\alpha=i/\hbar$, we note that
\beq
\partial(R^2\cdot\partial\S)=0, 
\label{continuequation}
\eeq
and 
\beq
{1\over2m}(\partial\S)^2=-{\hbar^2\over2m}{\Box(Re^{{i\over 
\hbar}\S})\over Re^{{i\over\hbar}\S}}+{\hbar^2\over2m}{\Box R\over R}. 
\label{ridentity2}
\eeq 
In analogy to the one dimensional identity we set 
\beq
\W_{rel}={\hbar^2\over2m}{\Box(Re^{{i\over\hbar}\S})\over Re^{{i\over\hbar}\S}}, 
\label{wrelativistic}
\eeq
that by (\ref{ridentity2}) implies 
\beq
Q_{rel}=-{\hbar^2\over2m}{\Box R\over R}. 
\label{ridentity3}
\eeq
Setting $\W_{rel}=1/2mc^2$ reproduces the Klein--Gordon equation and 
Eq. (\ref{ridentity2}) corresponds to the
Relativistic Quantum Hamilton--Jacobi Equation (RQHJE) \cite{bfm}. 
We further remark that the two particle case was considered in \cite{marco}.

\section{Phase space duality \& Legendre transformations}

We noted from eq. (\ref{mhje}) that the modified 
Hamilton--Jacobi equation allows for non--trivial solutions 
for the physical system with $W(q)\equiv0$. The 
QHJE therefore enables all physical states labelled 
by the potential function $W(q)$ to be connected 
to the trivial state via coordinate transformations, 
and facilitates the covariance of the QHJE. 
These properties are intimately related 
to phase space duality \cite{fm2}, which is implemented 
by the involutive nature of Legendre transformations.
Manifest phase--space duality may therefore 
provide the fundamental physical principle that is sought 
as the axiomatic principle for formulating quantum gravity.
In this respect we note that perturbative and nonperturbative
dualities play an important role in attempts to develop a 
fundamental understanding of string theory, with $T$--duality
being an important perturbative property of string theory \cite{gpr}.
$T$--duality in toroidal spaces exchanges momentum modes with winding
modes. We may therefore view $T$--duality as phase--space duality in
compact space. Furthermore, we may question whether $T$--duality
reflects a property of string theory which is also valid in non--toroidal
spaces. An additional important property of $T$ duality in
string theory is the existence of self--dual states under $T$--duality.

Manifest phase--space duality is implemented by the involutive nature 
of Legendre transformations. Recalling the relation between the 
momenta and coordinates via a generating function $p=\partial_q S$,
we define a dual relation via a new generating function $T(p)$ as
$p=\partial T$, where the generating functions are related 
by Legendre transformations as
\beq
S~=~p{\partial T\over\partial p}~-~T~~~~~~,~~~~~~~ %
T~=~q{\partial S\over\partial q}~-~S,
\label{legendreST}
\eeq
which in the stationary case reduces to 
\beq
S_0=p{{\partial T_0\over\partial p}}- T_0~~~~~~~~~~,~~~~~~~~~~%
T_0=q{{\partial S_0\over\partial q}}- S_0.
\label{legendreS0T0}
\eeq
Remarkably, the left--hand side of eq.
(\ref{legendreS0T0}) is invariant under 
M\"obius transformations 
$$q~\longrightarrow~ q^v~=~{Aq+B\over Cq+D},$$
with the induced transformations on $p$ and $T_0$
\begin{eqnarray}
p~  & &\longrightarrow~~~~~~ p^v~~~~ = ~~\rho^{-1}(Cq+D)^2p~~,~~~\rho~=AD-BC\nonumber\\
T_0~ & &\longrightarrow~~~ T_0^v(p^v)~ = ~T_0(p)~+~\rho^{-1}(ACq^2+2BCq+BD)p.\nonumber
\end{eqnarray}
We define a general coordinate transformation $q\rightarrow q^v=v(q)$
by the property that $S_0$ is a scalar function under $v$, {\it i.e.}
$S_0^v(q^v)=S_0(q)$. With each Legendre transformation we associate a second 
order differential equation \cite{fm2, fmijmp}, 
which in the stationary case is given by
\beq
\left({\partial^2\over\partial S_0^2}+U(S_0)\right)%
\left({{q\sqrt p}\atop \sqrt{p}}\right)~=0
\eeq
where $U(S_0)$ is given by the Schwarzian derivative of 
$q$ with respect to $S_0$,  
$$
U(S_0)={1\over2}\{q,S_0\}={1\over2}
\left(
{q^{\prime\prime\prime}\over q^{\prime}}-
{3\over2}\left({q^{\prime\prime}\over q^{\prime}}\right)^2
\right). 
$$
We may associate a second order differential equation with the 
involutive Legendre transformation. We therefore obtain 
manifest {$p\leftrightarrow q$} -- 
              {$S_0\leftrightarrow T_0$} duality with
\begin{eqnarray}
~~~~~p~&=&~{{\partial S_0}\over{\partial q}}~~~~~~~~~~~~~~~~~~~~~~~~~~~
~~~~~~~~~~q~=~{{\partial T_0}\over{\partial p}}\nonumber\\
~~~~~S_0~&=&~p{{\partial T_0\over\partial p}}- T_0~~~~~~~~~~~~~~~~~~~~~~~~~~~
T_0~=~q{{\partial S_0\over\partial q}}- S_0\nonumber
\end{eqnarray}
\begin{equation}
\left({{\partial^2~~} \over\partial S_0^2}+{U(S_0)}\right)
\left({{q\sqrt p}\atop \sqrt{p}}\right)=0
~~~~~~~~~~
\left({{\partial^2~~}\over\partial T_0^2}+{\cal V}(T_0)\right)
\left({{p\sqrt q}\atop \sqrt{q}}\right)=0\nonumber
\end{equation}
However, the crucial point is the existence of self--dual states,
with the property that $pq=\gamma=constant$, which are simultaneous 
solutions of the two pictures. In these cases $S_0=-T_0+constant$, and 
$$S_0(q)=\gamma\ln\gamma_qq~~~~~~~~~~~~~~~~~T_0(p)=\gamma\ln\gamma_p.p$$
Hence, $S_0~+~T_0~=~pq~=~\gamma$, where 
$\gamma_q\gamma_p\gamma={\rm e}$ and $\gamma_q$, $\gamma_p$ are constants. 
The remarkable point is that the self--dual states coincide with
the $W^0(q^0)\equiv 0$ states of the Quantum Hamilton--Jacobi Equation (QHJE), which 
render the consistency of the equivalence postulate for all physical 
states, and the compatibility of quantum mechanics with the underlying 
M\"obius symmetry of the QHJE. In the case of the Classical Hamilton--Jacobi
Equation (CHJE), it was noted that the solution in the case with
$W^0(q^0)\equiv 0$ correspond to $S_0=constant$, or more generally
$S_0=Aq^0+B$, with constants $A$ and $B$. The Legendre 
transformation is not defined for linear functions and therefore
classically in these case the phase--space duality would 
not be well defined. The existence of the self--dual quantum 
mechanical solutions, {\it i.e.} in the case with $W^{sd}=W^0=0$ 
and with $\gamma^{sd}=\pm\hbar /(2i)$, 
facilitate the consistency of 
phase--space duality for all physical states, 
as well as the consistency of the equivalence 
postulate, and compatibility with the underlying M\"obius
symmetry. It is crucial to appreciate that these properties
merely accommodate the basic quantum mechanical properties, 
and in that sense they are not esoteric at all. Namely, 
in this approach the emergence of $\hbar$ as the basic 
quantum mechanical parameter, arises as the covariantising
parameter in the QHJE, and enables the consistency of the formalism. 
Furthermore, it is noted that distinction between the 
classical and quantum mechanical cases in this approach
is primarily reflected in the distinction of the $W^0\equiv 0$ state. 
In this respect it will not be surprising if the formalism offers 
an intrinsic regularisation scheme. Furthermore, this 
intrinsic regularisation scheme is rooted in the global
M\"obius symmetry that underlies quantum mechanics. 
This is again not a surprise because the M\"obius 
symmetry implies that spatial space is compact. In turn
this implies the existence of a finite length scale in
the formalism and therefore an intrinsic regularisation 
scale. Therefore, a deeper understanding of the implications
of the M\"obius symmetry that underlies quantum mechanics,
brings forth the fertile soil on which the seeds of
quantum gravity may grow. 

\section{Intrinsic length scale}

The Schr\"odinger equation in the physical state with $W^0(q^0)=0$
is given by $${{\partial^2\Psi}\over{\partial q^2}}~=~0~,$$
and has two solutions $\psi_1=q^0$ and $\psi_2=constant$, which by the 
M\"obius symmetry must both be included in the formalism. The duality, 
manifested by the invariance under the M\"obius transformations,
therefore imply the existence of a length scale in the formalism. 
The corresponding solution of the QHJE is given by
\cite{fm2,fmijmp,fmpl}
\beq
{\rm e}^{{2i\over\hbar}S_0^0}={\rm e}^{i\alpha}
{{q^0+i{\bar\ell}_0}\over{q^0-i\ell_0}}, 
\label{w0solution}
\eeq
where the constant $\ell_0$ has the dimension of length
\cite{fmijmp,fmpl}, and the conjugate momentum takes the form
\beq
p_0=\partial_{q^0} S_0^0=
\pm{{\hbar (\ell_0+{\bar\ell}_0) }\over {2\vert q^0- i\ell_0\vert^2}}.
\label{pzerozero}
\eeq
It is seen that $p_0$ is null only when $q^0\rightarrow\pm\infty$. 
As we noted in section \ref{qhje} the condition that 
${\rm Re}\ell_0\ne 0$ is synonymous to the 
condition that $S_0\ne constant$, which 
is the basic quantum mechanics property in 
the equivalence postulate formalism. 
We can represent this nonvanishing length parameter as a 
combination of some fundamental constants in nature. {\it e.g.}
$\hbar$, $c$ and $G$. 
The requirement that 
$\lim_{\hbar\rightarrow0}p_0=0$ in the classical limit implies
that we can identify ${\rm Re}\,\ell_0$ with the 
Planck length, 
\cite{fmpl,fmijmp}
\beq
{\rm Re}\,\ell_0=\lambda_p= \sqrt{{\hbar G}\over c^3}, 
\label{setell0}
\eeq
The reason
being that this identification has the correct scaling properties
to reproduce the correct classical limit. Furthermore, 
we note from eq. (\ref{pzerozero})
that the condition
${\rm Re}\ell_0\ne 0$ serves as an ultraviolet cutoff, {\it i.e.}
$p_0$ is maximal for $q^0=-{\rm Im}\ell_0$, with
$$\hbox{Max} \vert p_0\vert ~=~ {\hbar\over {\hbox{Re} \ell_0}}.$$
Therefore, the consistency of the equivalence postulate 
formalism with the underlying M\"obius symmetry, implies
the existence of an intrinsic regularisation scale in 
quantum mechanics. Additionally, we may identify the 
quantum potential as an intrinsic curvature 
term of elementary particles. This provides further 
evidence that in this approach quantum 
mechanics regularises itself and a possible connection with 
theories of extended objects. This
is a mere reflection of the M\"obius symmetry and the 
compactness of spatial space. 

\section{Quantum potential as a curvature term}

Using the property of the Schwarzian derivative
$$\{S_0;q\}=-\left({{\partial S_0}\over {\partial q}}\right)^2\{q; S_0\},$$
we can rewrite the Quantum Stationary  Hamilton Jacobi Equation as,
$${1\over2m}\left({\partial S_0\over\partial\hat q}\right)^2+V(\hat q)-E=0,$$
where
$$\hat q=\int^q{dx\over\sqrt{1-{\hbar^2\over 2}\{q; S_0\}}}.$$
Flanders \cite{flanders} have shown that the Schwarzian 
derivative can be interpreted as a curvature term
of an equivalence problem for curves in ${\bf\rm P}^1$. 
Thus, the quantum potential, which is never vanishing, 
can be regarded as an intrinsic curvature term of elementary 
particle, and a deformation of the space geometry. 
Furthermore, we note from eq. (\ref{idwithvp})
that the quantum potential as a universal character, which 
distinguishes it from the gauge interactions that are 
dependent on the gauge charges. 
In higher dimensions the quantum
potential corresponds to the curvature of the 
function $R(q)$ as
$Q(q)\sim {{\Delta R(q)}/R}$. 
We can estimate the scale of the quantum potential \cite{de}.
In the case $W^0(q^0)\equiv0$ and using eq. (\ref{w0solution})
we obtain
\beq
Q^0~=~{\hbar^2\over4m}\{S_0^0,q^0\}~=~ - 
{{\hbar^2 ({\rm Re}\,\ell_0)^2}\over2m}
{1\over{\vert q^0-i\ell_0\vert^4}}.
\label{q0}
\eeq
Taking $m\sim100GeV$; ${\rm Re}\ell_0=\lambda_p\approx
10^{-35}m$ and $q^0$ as the size of the observable universe
$q^0\sim93{\rm Ly}$, gives $\vert Q\vert\sim 10^{-202}eV$. 
The expected contribution of the quantum
potential to the vacuum energy is very small. 
Nevertheless, we see from eq. (\ref{q0}) that the 
classical limit $Q(q)\rightarrow0$ in fact correspond to the
decompactification limit $q^0\rightarrow\infty$.

\section{Conclusions}

The observation of a scalar resonance at the the LHC reinforces
the picture that the Standard Model provides a viable effective 
description of all subatomic data up to the Planck scale.
The synthesis of gravity with quantum mechanics remains 
an open problem. An alternative approach to quantising general relativity 
is to formulate a geometric approach to quantum mechanics. This
is precisely the aim in the equivalence postulate approach 
to quantum mechanics. What is revealed is the key role 
of the M\"obius symmetry that underlies quantum mechanics
in this formalism. In turn the M\"obius symmetry provides
the key to a proper understanding of the geometry of the 
quantum spacetime. In this respect the formalism is 
intimately related to phase space duality manifested 
by the involutive nature of Legendre transformations. 
Recalling their role in thermodynamics we may envision 
that the Legendre transformations merely transfer 
from one set of variables to another, and neither
set should be thought of as more fundamental. In this context
the ubiquity of the variable themselves allows us to 
consider transformations between the space coordinates
and the wave function itself \cite{fm1}, without considering
one as being primary and the other secondary. Their only merit
is their usefulness for a particular physical measurement. 
In this respect possible observational signatures of
the M\"obius symmetry underlying quantum mechanics
may exist in the microwave background radiation \cite{cmb}. 
Additionally, the quantum potential may lead to modified 
dispersion relations \cite{opera} with possible observational 
consequences \cite{subir}.

\medskip
{\bf Acknowledgements}

This work is supported in part by the STFC under contract ST/G00062X/1.

\end{document}